\documentclass[aps,amsfonts,amsmath,prd,preprint,nofootinbib]{revtex4-1}
\newcommand{\beq}{\begin{equation}}
\newcommand{\eeq}{\end{equation}}
\usepackage{epsfig}

\begin{document}

\title{Transdimensional Tunneling in the Multiverse}

\author{Jose J. Blanco-Pillado, Delia Schwartz-Perlov, and Alexander Vilenkin}

\affiliation{
Institute of Cosmology, Department of Physics and Astronomy,\\ 
Tufts University, Medford, MA 02155, USA
}

\begin{abstract}

Topology-changing transitions between vacua of different effective 
dimensionality are studied in the context of a 6-dimensional 
Einstein-Maxwell theory. The landscape of this theory includes a 
$6d$ de Sitter vacuum ($dS_6$), a number of $dS_4 \times S_2$ 
and $AdS_4 \times S_2$ vacua, and a number of $AdS_2 \times S_4$ vacua.
We find that compactification transitions $dS_6 \to AdS_2 \times S_4$ 
occur through the nucleation of electrically charged black hole pairs, 
and transitions from $dS_6$ to $dS_4 \times S_2$ and $AdS_4 \times
S_2$ occur through the nucleation of magnetically charged spherical 
black branes. We identify the appropriate instantons and describe 
the spacetime structure resulting from brane nucleation.

\end{abstract}

\maketitle

\section{Introduction}

String theory, as well as other higher-dimensional theories, suggests
the existence of a landscape of vacua with diverse properties.  The
vacua are characterized by different compactifications of extra
dimensions, by branes wrapped around extra dimensions in different
ways, by different values of fluxes, etc.  The number of
possibilities is combinatorial and can be as high as $10^{1000}$
\cite{Susskind}.  In the cosmological context, high-energy vacua drive
exponential inflationary expansion of the universe.  Transitions
between different vacua occur through tunneling \cite{CdL} and quantum
diffusion \cite{AV83,Linde86}, with regions of different vacua
nucleating and expanding in the never-ending process of eternal
inflation.  As a result, the entire landscape of vacua is explored.
This ``multiverse'' picture has been a subject of much recent
research.

Most of this recent work has focused on the sector of the landscape
with 3 large spatial dimensions and the rest of the dimensions
compactified.  If all vacua had this property, we would be able to
describe the entire multiverse by a 4-dimensional effective
theory.  However, the landscape generally includes states with
different numbers of compactified dimensions.  Even a simple toy model
based on a 6-dimensional Einstein-Maxwell theory exhibits vacua
with 0, 2, and 4 compact dimensions.  We therefore
expect the multiverse to include regions of different (effective)
dimensionality \cite{Linde88}.

Topology-changing transitions between vacua with different numbers of
compact dimensions have been discussed in
Refs.~\cite{Linde88,Decompactification,BPSPV09,Carroll09} in the
context of Einstein-Maxwell theory compactified on a sphere. The
radius of the compact dimensions in this theory is stabilized by a
magnetic flux through the sphere, but the resulting vacua are only
metastable and can decay by quantum tunneling through a
barrier.\footnote{The model of Ref.~\cite{Linde88} had an extra
ingredient -- the inflaton field -- and a different mechanism of
decompactification. The inflaton undergoes quantum diffusion in the
course of eternal inflation, and the compact dimensions are
destabilized whenever the inflaton potential gets below certain
critical values.}  The tunneling results in the formation of an
expanding bubble, inside of which the compact sphere is destabilized
and its radius grows with time.  Transitions of this sort are called
{\it decompactification} transitions.  The {\it compactification}
tunneling transitions go in the opposite direction: the number of
compact dimensions increases from parent to daughter vacuum.  Such
transitions have been discussed in the interesting recent paper by
Carroll, Johnson and Randall \cite{Carroll09}, who also obtained an
estimate of the transition rate.  Our main goal in the present paper
is to elucidate the nature of the tunneling instantons and the
spacetime geometry resulting from topology-changing transitions.

In the next Section we begin with a brief review of flux vacua in the
$6d$ Einstein-Maxwell theory.  Compactification tunneling transitions 
from $6d$ to effectively $2d$ and $4d$ spacetimes are discussed in 
Sections III and IV, respectively, and decompactification transitions 
are discussed in Section V. Our conclusions are summarized and 
discussed in Section VI.

\section{The landscape of $6d$ Einstein-Maxwell theory} \label{$6d$ landscape}

We shall consider a 6-dimensional Einstein-Maxwell theory, 
\beq
S=\int{d^6 {x} \sqrt{-\tilde g} \left({1\over 2} {\tilde R}^{(6)} 
- {1\over 4} F_{MN} F^{MN} - {\Lambda_6}\right)},
\label{EM-6D-action}
\eeq 
where $M,N = 0...5$ label the six-dimensional coordinates, $F_{MN}$ is
the Maxwell field strength, ${\Lambda_6}$ is the six-dimensional 
cosmological constant, ${\tilde R}^{(6)}$ is the 6-dimensional
curvature scalar, and we use units in which the $6d$ Planck mass is $M_6 = 1$.
We shall assume that $\Lambda_6>0$.  This model has a
long pedigry \cite{Freund-Rubin,EM6D}; more recently it has been
discussed as a toy model for string theory compactification
\cite{flux-compactifications}. 

Flux vacua are described by solutions of this model with the spacetime
metric given by a $(6-q)$-dimensional maximally symmetric space of
constant curvature and a static $q$-dimensional sphere of
fixed radius, namely a metric of the form,
\beq
ds^2= {\tilde g}_{MN} dx^M dx^N = {\tilde g}_{\mu \nu} d x^{\mu}
d x^{\nu} + R^2 d\Omega_q^2~.
\label{6D-metric}
\eeq 
Here, ${\tilde g}_{\mu\nu}$ is the metric of the $(6-q)$-dimensional
de Sitter, Minkowski, or anti-de Sitter space, and $d\Omega_q^2$ is
the metric on a unit $q$-sphere.  We shall use the convention that the
Greek indices label the large dimensions and take values
$\mu,\nu=0,1,...,5-q$. The compact dimensions will be labeled by
indices from the beginning of the Latin alphabet, $a,b=6-q,...,5.$

The simplest solution of the model, corresponding to $q=0$, is the
$6d$ de Sitter space with $F_{MN}=0$.

For $q=2$, the only ansatz for the Maxwell field that is
consistent with the symmetries of the metric is a monopole-like
configuration on the extra-dimensional 2-sphere \cite{EM6D},
\beq
F_{ab}= {Q_m\over{4\pi}} \sqrt{g_2} \epsilon_{ab},
\label{Fab}
\eeq 
where $Q_m$ is the corresponding magnetic charge and $g_2$ is the
determinant of the metric on a unit 2-sphere. All other
components of $F_{MN}$ are equal to zero.  We shall assume that our
model includes electrically charged particles with elementary charge $e$
(not shown in the action (\ref{EM-6D-action})).  Then the magnetic
charge is subject to the usual charge quantization condition,
\beq
Q_m={2\pi n\over{e}} ,
\label{Q}
\eeq
where $n$ is an integer.  This sector
of the model can be described in terms of an effective $4d$ field
theory.  Representing the $6d$ metric as
\beq
ds^2= e^{- \psi(x)/M_4} g_{\mu \nu}
dx^{\mu} dx^{\nu} + e^{\psi(x)/M_4} R^{2}~d\Omega_2^2
\label{psimetric}
\eeq
and integrating over the internal manifold, we obtain
\beq
S= \int{d^4 x \sqrt{-g}\left({1\over 2} M_4^2 R^{(4)} - {1\over 2}
  \partial_{\mu} \psi \partial^{\mu} \psi - V(\psi)\right)}.
\label{4Deffectiveaction}
\eeq
Here, the potential for the size of the internal dimension
is
\beq
V(\psi)= 4\pi \left({{n^2}\over{8 e^2 R^2}}
e^{-3\psi/M_4} - e^{-2\psi/M_4} + {R^2 \Lambda_6} e^{-\psi/M_4}
\right)
\label{V} 
\eeq
and 
\beq
M_4^2 = 4 \pi R^2 ,
\label{M4}
\eeq
is the $4d$ Planck mass.

For any particular value of $n$, we can set
the minimum of the potential to be at $\psi =0$, by setting
\beq
R^2 = {1\over{\Lambda_6}} \left(1 - \sqrt{1 - {{3 n^2
\Lambda_6}\over {8 e^2}}}\right).
\label{R2}
\eeq
The $4d$ cosmological constant is the value of the potential at 
this minimum and is given by
\beq
\Lambda_4 = 
V(\psi=0,n)={{4\pi}\over {3}} \left(1- 2 \sqrt{1 - {{3 n^2
\Lambda_6}\over {8 e^2}}}\right)~.
\label{Lambda4}
\eeq
Positive-energy vacua are obtained for
\beq
n>n_0 \equiv \left({{2}\over{\Lambda_6}}\right)^{1/2} e .
\eeq

\begin{figure}
\centering\leavevmode
\epsfysize=8cm \epsfbox{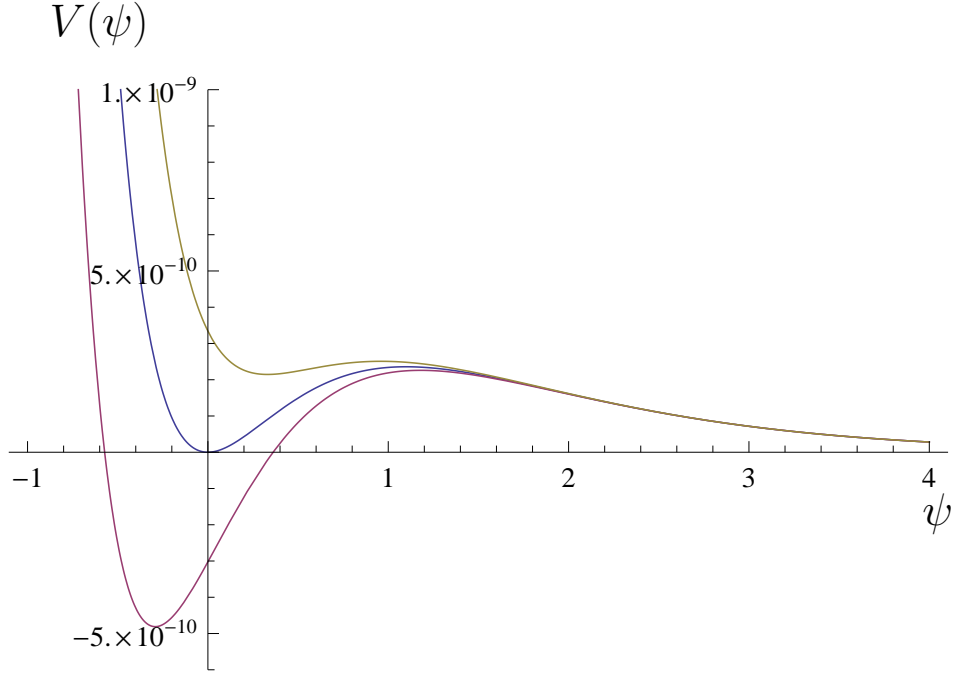}
\put(-10,55){\Large {\bf {$\psi$}}}
\put(-340,240){\Large {\bf {$V(\psi)$}}}
\caption {Plot of the $4d$ effective potential, in units of $M_4^4$,
as a function of the field $\psi$. We show the potential for 3
different values of the flux quantum $n = 180, 200, 220$. The rest of
the parameters of the model are fixed to $e^2 = 2$ and 
$\Lambda_6 = 10^{-4}$, so $n_0 =200$.}
\label{potential}
\end{figure}

The potential also has a maximum at some $\psi > 0$ and approaches
zero at $\psi\to\infty$, as illustrated in Fig.~\ref{potential}. It is clear from the figure that a positive-energy vacuum can decay by tunneling through a barrier, leading to decompactification of extra dimensions\footnote{Perturbative stability in similar models has been discussed in \cite{Bousso:2002fi,Krishnan:2005su}.}.

No vacuum solutions exist for $n > n_{max}=2n_0/\sqrt{3}$.  
The corresponding value of $Q_m$ is 
\beq
Q_{max}^{(m)}=4\pi \left({{2}\over {3\Lambda_6}}\right)^{1/2}.
\label{Qmax(m)}
\eeq
A large landscape is possible only if $n_0\gg 1$, or equivalently
  $\Lambda_6\ll e^2$.

Finally, in the $q=4$ sector the Maxwell field takes the form
\beq
F_{\mu\nu} = {Q_e \over{A_4 R^4}}\sqrt{-g_2}\epsilon_{\mu\nu}.
\eeq
where $R$ is the radius of the 4-sphere compactification 
manifold, $A_4=8\pi^2/3$ is the 4-volume of a unit 4-sphere,
and $\sqrt{-g_2} = \sqrt{-\tilde g_{\mu \nu}}$ is the
determinant of the metric in the 2 large dimensions.
The charge ${Q}_e$ is quantized in terms of the 
elementary charge $e$ by the relation
\beq
Q_e = n e .
\label{Qe}
\eeq
One can also represent this sector as a flux compactification model 
in the dual formulation in terms of a 4-form flux. This is discussed 
in Appendix  \ref{appelecsec}.

We can now look for solutions of our model where the  
spacetime described by the large 2 dimensions is characterized by a
constant scalar curvature, $R^{(2)}= 2H^2$. Einstein's equations 
then reduce to the following relations: 
\beq 
H^2 = {\Lambda_6 \over {2}}\left(1- {3{Q_e^2}\over{2
    A_4^2\Lambda_6 R^8 }}\right) , 
\label{H2}
\eeq 
\beq 
{3 \over {R^2}} = {\Lambda_6
  \over {2}}\left(1+{{Q_e^2}\over{2 A_4^2\Lambda_6 R^8}} \right) .  
\label{R^2}
\eeq
Solutions of these equations are discussed in Appendix
\ref{quarticequation}. For $n>n_{max}=(9A_4/e)(3/2\Lambda_6)^{3/2}$,
there are no solutions. For $n<n_{max}$, there are two solutions, 
one with a positive and the other with a negative value of $H^2$. 
(This situation is similar to that in the $q=2$ sector, where 
we also have two solutions, one
corresponding to a minimum and the other to a maximum of the
potential.  In both cases, the higher-energy solutions are unstable;
see the discussion in Sec.III).   As $n$ is increased, the two solutions
approach one another, until they both reach $H=0$ at $n=n_{max}$.
The corresponding maximum value of $Q_e$ is given by
\beq
Q_{max}^{(e)}=24\pi^2 \left({{3}\over {2\Lambda_6}}\right)^{3/2}~.
\label{Qmax(e)}
\eeq

Once again, we can obtain an effective $2d$ theory by integrating 
out the extra dimensions.  The resulting action is that of a $2d$ 
dilatonic gravity.  The action
cannot be reduced to the form (\ref{4Deffectiveaction}), with Einstein
gravity plus a minimally coupled scalar, essentially because the
Einstein Lagrangian is a pure divergence in $2d$.  Since the ``dilaton"
field $\psi$ has a non-minimal coupling to gravity, the radius of the
compact dimensions cannot be found by simply finding the extrema of
the dilaton potential but one can nevertheless identify the
solutions presented above as solutions of the scalar tensor theory in
$1+1$ dimensions.

To summarize, we have found that the landscape of vacua in our theory
includes a $dS_6$ vacuum, a number of $dS_4$ and $AdS_4$ vacua with
extra dimensions compactified on $S_2$, and a number of $AdS_2$ vacua
with extra dimensions compactified on $S_4$.  We expect that quantum
transitions should occur from $dS_6$ and $dS_4\times S_2$ vacua to all other vacua, resulting
in an eternally inflating multiverse.  The AdS vacua are terminal, in
the sense that all AdS regions collapse to a big crunch singularity.

Transitions between $dS_4 \times S_2$ vacua with different flux
quantum numbers $n$ were discussed in Ref.~\cite{BPSPV09}, where it
was shown that such transitions proceed through nucleation of
magnetically charged branes.  The theory does have black brane
solutions with the necessary properties \cite{Gibbons95,Gregory}.  Our
focus in the present paper is on the topology-changing transitions.
As we shall see, such transitions also involve nucleation of charged
black branes or black holes.

\section{From $dS_6$ to $(A)dS_2\times S_4$}\label{blackhole}

\subsection{The instanton}

As discussed in Ref.~\cite{Carroll09}, transitions from
 $dS_6$ to $AdS_2 \times S_4$ can be
mediated by nucleation of black holes in $dS_6$.  The corresponding
instanton is the Euclideanized $6d$ Reissner-Nordstrom-de Sitter (RNdS)
solution \cite{Tangherlini},
\beq
ds^2 =  f(r)d\tau^2 + f(r)^{-1}dr^2 + r^2 d\Omega_4^2,
\label{BHinstanton}
\eeq
where
\beq
f(r)= 1 - {2{\tilde M}\over{r^3}}+{{\tilde Q}^2\over{r^6}} -H_6^2 r^2 .
\eeq
Here, 
\beq
H_6 = \sqrt{{{\Lambda_6}\over {10}}}
\eeq
 is the expansion rate of $dS_6$, ${\tilde M}$  is the mass parameter,
 and ${\tilde Q}$ is related to the black hole charge $Q_e$ as 
\beq
Q_e=\sqrt{12}A_4{\tilde Q}.  
\label{qtilde}
\eeq
The electric field is given by
\beq
F_{\tau r}={iQ_e\over{A_4 r^4}} ,
\eeq
and $Q_e$ is quantized in units of the elementary
charge, as in Eq.~(\ref{Qe}).\footnote{This is based on the definition
of the charge as the integral of the flux through the $S_4$ sphere 
around the pointlike charged object\beq
Q = \int{F_{\tau r}~ r^4~\sin^3{\theta}\sin^2{\phi}\sin{\psi} d\theta
  d\phi d\psi}
\eeq
}

Zeros of the function $f(r)$ correspond to horizons in the Lorentzian
solution.  There are generally three horizons: the inner ($r_-$) and
outer ($r_+$) black hole horizons and the cosmological horizon $r_c$.
The range of the variable $r$ in (\ref{BHinstanton}) is 
\beq 
r_+ \leq r \leq r_c,
\label{range}
\eeq
so that the metric is positive-definite and non-singular, with
\beq
f(r_+)=f(r_c)=0 .
\label{f=0}
\eeq  
The Euclidean time $\tau$ is a cyclic variable with a period chosen to
eliminate conical singularities at $r=r_+,r_c$.

Nucleation of charged black holes in dS space has been extensively
studied, both in $4d$ \cite{Moss,Romans,Mann} and in higher dimensions
\cite{Dias}. For relatively small values of ${\tilde Q}$,
 the relevant instantons have the topology of $S_2 \times S_4$, 
and the avoidance of a conical singularity in the geometry imposes the condition 

\beq
|f'(r_+)|=|f'(r_c)|. 
\label{lukewarm}
\eeq
The period of $\tau$ is then
\beq
\Delta\tau = {4\pi\over{|f'(r_c)|}}.
\eeq
These are the so-called ``lukewarm instantons".

\begin{figure}
\centering\leavevmode
\epsfysize=6cm \epsfbox{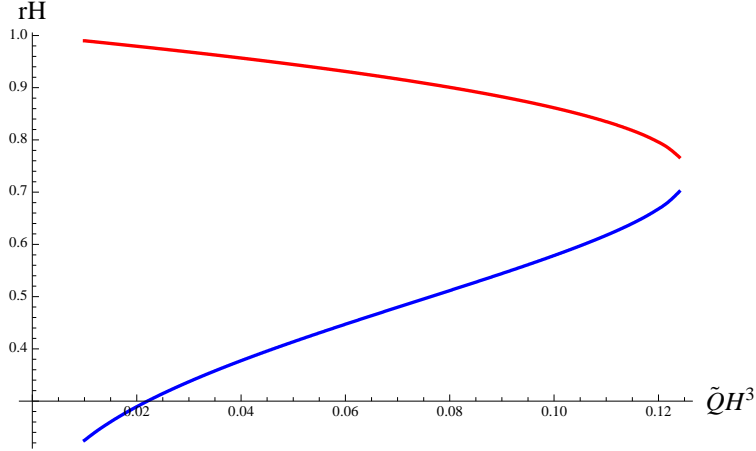}
\caption {Plot of  $r_+ H_6$ (bottom) and $r_c H_6$ (top) vs. ${\tilde
    Q}H^3_6$.  }
\label{horizonradii}
\end{figure}

The condition (\ref{lukewarm}) implies that the black hole and
cosmological horizons have the same temperature and imposes a relation
between the parameters $\tilde M$, $\tilde Q$, and $H_6$.  In our
case, $H_6$ and $e$ should be regarded as fixed by the model, while
$\tilde M$ is an adjustable parameter.  The horizon radii $r_+$
and $r_c$ are plotted in Fig.~\ref{horizonradii} as functions of the 
charge ${\tilde Q}$.  (More precisely, we plot $r_+ H_6$ 
and $r_c H_6$ vs. ${\tilde Q}H^3_6$.)
As ${\tilde Q}$ is increased at a fixed $H_6$, the black hole horizon
radius $r_+$ grows until it coincides, at 
${\tilde Q} \approx 0.125 H_6^{-3} \equiv {\tilde Q}_c$, with the 
cosmological horizon $r_c$.

In the limit when the black hole is much
smaller than the cosmological horizon,
\beq
r_+ \ll r_c ,
\label{smallBH}
\eeq
it follows from Eqs.~(\ref{f=0}), (\ref{lukewarm}) that,  up to the
leading corrections in ${\tilde q}\equiv {\tilde Q}H_6^3$, we have 
\beq
{\tilde M}\approx {\tilde Q}\left(1-{1\over{2}}{\tilde q}^{2/3}\right),
\label{MQ}
\eeq
\beq
r_+\approx {\tilde Q}^{1/3}\left(1+{1\over{3}}{\tilde q}\right)^{1/3},
\label{r+Q}
\eeq
\beq
r_c\approx H_6^{-1} (1-{\tilde q}),
\label{rcQ}
\eeq 
and 
\beq
\Delta\tau\approx 2\pi H_6^{-1}(1+4{\tilde q}).
\label{tauQ}
\eeq
Thus, the instanton in this limit
describes the nucleation of nearly extremal black holes in $dS_6$.  These
black holes can be thought of as 0-branes in our theory.
The mass $M$ of the black holes can be
determined from the behavior of the metric in the range $r_+\ll r \ll r_c$.
This gives \cite{MyersPerry86}
\beq
M={32\pi^2\over{3}}{\tilde M} .\label{ADMmass}
\eeq

\begin{figure}
\centering\leavevmode
\epsfysize=8cm 
\epsfig{file=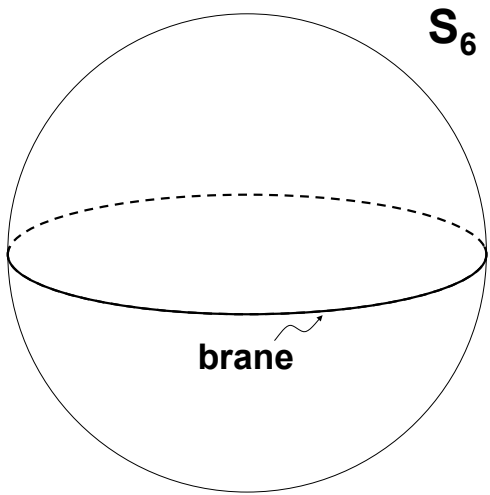,width=8cm,angle=-90}
\caption {Topology-change instanton. The charged black hole worldline 
runs along a big circle on a 6-sphere.  The spherical geometry is
distorted in the vicinity of the worldline.}
\label{Alexfig1}
\end{figure}

Geometrically, we can picture the RNdS instanton in the following way.
Consider the surface $r=r_*$ with $r_+\ll r_*\ll r_c$.  This is a
$5d$ surface with topology $S_4 \times S_1$.  It encircles the black
hole horizon $r=r_+$ with a sphere of radius $r_* \gg r_+$, so that
$f(r_*)\approx 1$.  Outside of this surface, the metric is that of a
6-sphere, which is the Euclideanized $dS_6$.  The $S_1$ comes from the
periodic time direction.  The length of this circle is given by the
period $\Delta\tau$ and is approximately the length of a big circle on
$S_6$.  Thus, the instanton can be pictured as a 0-brane whose
worldline runs along a big circle on $S_6$, as illustrated in Fig.~\ref{Alexfig1}.

Instantons of this form are known to describe the production of
  particle-antiparticle pairs in de Sitter space \cite{Basu}.  
The corresponding instanton action can be estimated as
\beq
S_{inst}\approx -{16\pi^3 \over{3H_6^4}} + {2\pi M\over{H_6}} ,
\label{BHaction}
\eeq
where the first term is the $6d$ Euclidean de Sitter action,
$S_{dS_6}$, and $M$ is the particle mass.  The second term on the
right-hand side of (\ref{BHaction}) is the contribution of the
particle worldline action,
\beq
M\int d\tau~.
\eeq
One might also expect to see $O(M)$ corrections to the de Sitter
action of a comparable magnitude, caused by distortions of the de
Sitter geometry due to the presence of the mass.  However, since the
action is minimized on the de Sitter solution, the linear in $M$
correction to the action should vanish in the limit of small $M$.

In our case, the instanton describes nucleation of a pair of
oppositely charged black holes with an initial separation of
$2H_6^{-1}$.  The black holes are then driven apart by the de Sitter
expansion.  We expect that for ${\tilde q} \ll 1$ the action can
be approximated by Eq.~(\ref{BHaction}). Disregarding the
pre-exponential factor, the nucleation rate is then given by
\beq
\Gamma \sim \exp (S_{dS_6}-S_{inst}) \sim \exp(-2\pi M/H_6) \sim \exp
\left({4\pi\over{\sqrt{3}}}{Q_e\over{H_6}}\right) ,
\eeq
where in the last step we have used the relations (\ref{ADMmass}), 
$\tilde M \approx \tilde Q$, and (\ref{qtilde}). This agrees with 
the intuitive expectation that $\Gamma \sim \exp(-M/T_{dS})$, 
where $T_{dS}=H_6/2\pi$ is the Gibbons-Hawking temperature of 
$dS_6$.\footnote{The nucleation rate can be modified if the 
action includes a topological term, $S_{top}=-\alpha\chi$, 
where $\alpha=const$ and $\chi$ is the Euler character \cite{Parikh}. 
In our case, $\chi(S_6)=2$ and $\chi(S_4\times S_2)=4$, so the
additional factor in $\Gamma$ is $e^{2\alpha}$. The same factor would 
also appear in the nucleation rate of black branes discussed in Section IV.A.}   

An alternative tunneling channel is given by the so-called charged
Nariai instanton, which describes nucleation of maximally large black
holes having $r_+ = r_c$ \cite{Bousso1,Bousso2}. This instanton has a
simple geometry in the form of a product of two round spheres,
$S_2\times S_4$, and is a Euclidean continuation of the $dS_2 \times
S_4$ vacuum solution that we discussed in Section II. 

\begin{figure}
\centering\leavevmode
\epsfysize=6cm \epsfbox{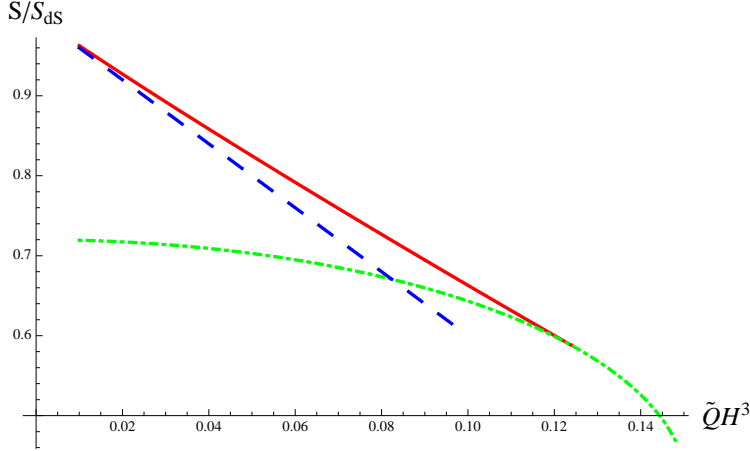}
\caption {Plot of  $S/S_{dS}$ vs. ${\tilde Q}H^3_6$ for the lukewarm 
(red/solid line), Nariai (green/dashed-dotted line) and approximate 
lukewarm (blue/dashed line) actions.}
\label{actions}
\end{figure}

The action for lukewarm and charged Nariai instantons in an arbitrary
number of dimensions has been calculated in \cite{Dias}.  In the
6-dimensional case we get\footnote{Note that this expression has been
  corrected in the latest electronic version of the paper in
  \cite{Dias}. We would like to thank Oscar Dias for clarifying this
  point to us.}
\beq
S_{lukewarm}=16\pi^2\Delta\tau \left[-\frac{1}{6}H_6^2 (r_c^5 -
  r_+^5) + \frac{1}{2}{\tilde Q}^2(r_+^{-3} - r_c^{-3})\right]
\label{Sluke}
\eeq
and
\beq
S_{Nariai}= -\frac{32\pi^3}{3}R^4.
\label{SNariai}
\eeq
where $R$ is the radius of the compact 4-sphere, which can be
determined from Eq.~(\ref{R^2}).  The actions (\ref{Sluke}), (\ref{SNariai}),
in units of the de Sitter action $S_{dS_6}$, are plotted in Fig.~\ref{actions} as 
functions of ${\tilde q}={\tilde Q} H_6^3$. The lukewarm instanton exists only for
${\tilde Q}<{\tilde Q}_c$, and in all this range its action is smaller\footnote{Note that $S_{lukewarm}$ and $S_{Nariai}$  are negative  (in addition to  $S_{dS_6} < 0$), and that is why in Fig. \ref{actions}  the graph for the lukewarm action is above that for the Nariai action.} (and thus the
nucleation rate is higher) than that for the charged Nariai instanton.
As ${\tilde Q}$ approaches ${\tilde Q}_c$ from below, the two 
instanton actions approach one another and coincide at ${\tilde
  Q}={\tilde Q}_c$. The Nariai instanton is the only
relevant tunneling channel in the range ${\tilde Q}_c < {\tilde Q} <
{\tilde Q}_{max}$,\footnote{There is yet another kind of instanton,
the so-called ``cold" instanton, describing nucleation of extremal
Reissner-Nordstrom black holes. Such instantons do not seem to
describe compactification transitions, so we do not consider them
here.}  where ${\tilde Q}_{max}= Q_{max}^{(e)}/\sqrt{12}A_4$ and 
$Q_{max}^{(e)}$, given by Eq.~(\ref{Qmax(e)}), is the largest value 
of $Q_e$ above which no instanton solutions exist.

Also shown in Fig.~\ref{actions} is the approximate action (\ref{BHaction}).  As
expected, it is in a good agreement with the exact action at small
values of ${\tilde Q}$. We have also verified this directly, by expanding the
lukewarm instanton action (\ref{Sluke}) at small ${\tilde Q}$. This gives
\beq
S_{lukewarm}=-{16\pi^3\over{3H_6^4}}(1-4{\tilde q}+{\tilde q}^{4/3} +...),
\eeq
where the neglected terms are $O({\tilde q}^{5/3})$. The first two
terms in this expansion coincide with our approximate formula (\ref{BHaction}).

\subsection{Spacetime structure}

To analyze the spacetime structure resulting from the tunneling, we
use a Lorentzian continuation of the metric (\ref{BHinstanton}),
$\tau\to it$.  Introducing a new radial variable $\xi$ as
\beq
f(r)^{-1/2}dr = d\xi,
\eeq
we have
\beq
ds^2 = -h(\xi)dt^2 + d\xi^2 + r(\xi)^2 d\Omega_4^2,
\label{BHmetric}
\eeq
where $h(\xi)\equiv f(r(\xi))$.  We can choose the origin of $\xi$ so
that $\xi=0$ at $r=r_+$.  Then, in the vicinity of $r= r_+$ we have
\beq
h(\xi)\approx \xi^2.
\eeq
We can now continue the metric across the black hole horizon by
replacing $\xi\to it$, $t\to \xi$,\footnote{Here we follow the
  discussion in \cite{Carroll09}.} 
\beq
ds^2 = -dt^2 + t^2 d\xi^2 + r_+^2 d\Omega_4^2.
\label{nearhorizon}
\eeq
This describes an expanding $2d$ FRW universe with 4 extra dimensions
compactified on a sphere.  The big bang at $t=0$ is non-singular and
corresponds to the horizon at $r=r_+$.   

The form (\ref{nearhorizon}) applies only in the vicinity of $t=0$.
More generally, the metric behind the horizon can be expressed as 
\beq
ds^2 = -dt^2 + a^2(t) d\xi^2 + r^2(t) d\Omega_4^2,
\label{behindhorizon}
\eeq
with 
\beq
F_{\mu\nu}= Q_e {{a(t)}\over {A_4 r(t)^4}} \epsilon_{\mu\nu} ~~~~ (\mu,\nu = 0,1).
\eeq

In the limit (\ref{smallBH}), and at $t\gg r_+$, we expect the metric
to approach that of $AdS_2\times S_4$.  The
extra dimensions in this metric are stabilized by the electric field;
that is why the black holes mediating the topology change need to
be charged. 

\begin{figure}
\centering\leavevmode
\epsfig{file=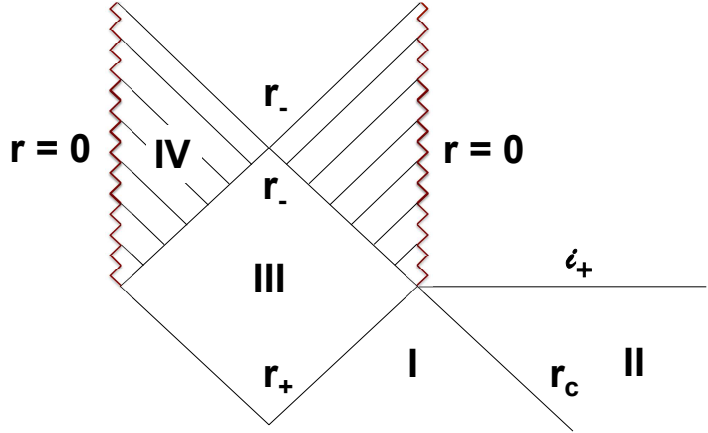,width=8cm,angle=-90}
\caption {Penrose diagram for a charged black hole mediating a 
topology-changing transition $dS_6\to AdS_2\times S_4$ .
Perturbations result in a singularity at $r=r_-$, so the striped 
regions are not accessible.}
\label{Alexfig2}
\end{figure}

\begin{figure}
\centering\leavevmode
\epsfig{file=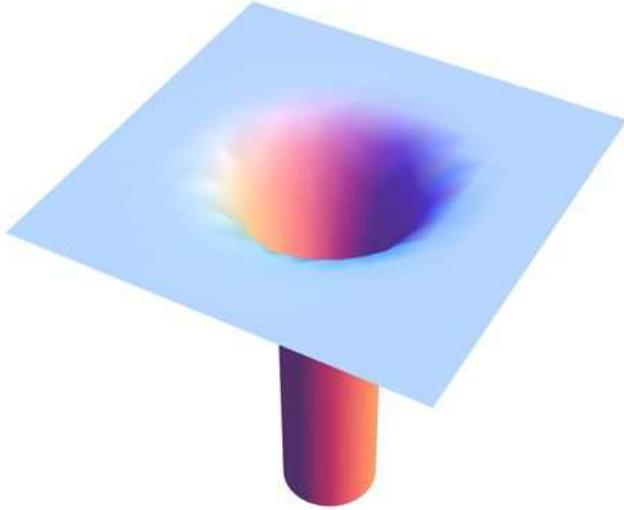,width=8cm,angle=-90}
\caption {A spacelike slice through a lower-dimensional analogue of 
the black hole topology-changing region. In this funnel-like geometry, 
the flat region at the top has two large dimensions, while the tube 
of the funnel has one large and one compact dimension.}
\label{funnel-picture}
\end{figure}

The Penrose diagram for a RNdS black hole is shown in Fig.~\ref{Alexfig2}.  Region
I in the diagram is the region $r_+ < r < r_c$ covered by the static
coordinates (\ref{BHmetric}).  Region II is the exterior
asymptotically de Sitter space, and $i_+$ is its spacelike future
infinity.  Region III is the part of the spacetime behind the black
hole horizon, which is described by the metric (\ref{behindhorizon}).
It corresponds approximately to $AdS_2 \times S_4$, with the horizon
at $r_+$ playing the role of the big bang and that at $r_-$ the role
of the big crunch of the AdS space.  In a pure AdS space, both
horizons are non-singular, so the black hole is a traversable
wormhole, and it is possible for a timelike curve to go from region I
across region III into a region similar to I and into another $dS_6$
space, or to a timelike singularity in region IV.  However, if
perturbations are included, the horizon at $r=r_-$ develops a true
curvature singularity, and the metric cannot be extended beyond region
III. A spacelike slice through the topology-changing region is
illustrated in a lower-dimensional analogue in
Fig. \ref{funnel-picture}.

Thus, we conclude that, from the viewpoint of a $6d$ observer, the
topology-changing transitions $dS_6\to AdS_2\times S_4$ look like
nucleations of pairs of electrically charged black holes.  Each black
hole contains an infinite $AdS_2\times S_4$ universe behind its
horizon.  

The Lorentzian continuation of the charged Nariai instanton 
is a $dS_2\times S_4$ solution, corresponding to one of the positive-energy 
$(H^2 >0)$ $q=4$ vacua that we discussed in Section II. These 
solutions are known to be unstable: small perturbations cause them 
to disintegrate into $AdS_2\times S_4$ and $dS_6$ regions \cite{Bousso2}.

\section{From $dS_6$ to $(A)dS_4 \times S_2$}

\subsection{The instanton}

We shall now argue that topology change from $dS_6$ to $dS_4\times
S_2$ and to $AdS_4 \times S_2$ can occur through the nucleation of
spherical black 2-branes.  Since the radius of $S_2$ is stabilized by
magnetic flux, the branes have to be magnetically charged.

It is well known that Einstein-Maxwell theory with $\Lambda_6 = 0$ has
magnetically charged, asymptotically flat black brane solutions
\cite{Gibbons95, Gregory}.  In particular, the extreme 2-brane in $6d$ is
described by the metric,  
\beq  
ds^2 = \left( 1-{r_0\over{r}}\right)^{2/3} (-dt^2 + dx^2 + dy^2) +
\left(1-{r_0\over{r}} \right)^{-2} dr^2 + r^2 d\Omega_2^2, 
\label{flatbrane}
\eeq
with
\beq
r_0 = {\sqrt{3} Q_m \over{8\pi}}.
\label{r0}
\eeq
The surface $r=r_0$ is a non-singular event horizon.  The parameter
$Q_m$ in (\ref{r0}) is the quantized magnetic charge (\ref{Q}). The mass 
per unit area is equal to the brane tension and is given by \cite{Lu,BPSPV09}
\beq
 T_2 = {2Q_m\over{\sqrt{3}}}.
\label{T2}
\eeq 
Note that the metric (\ref{flatbrane}) is invariant with respect to Lorentz 
boosts in the  $x$ and $y$ directions, as expected.  Using the results 
of Refs.~\cite{Gregory} and \cite{Gibbons95}, it can be shown that 
(\ref{flatbrane}) is the only solution which is both non-singular and 
boost-invariant.

Generalizations of black brane solutions to asymptotically de Sitter
background are not known analytically, but can be found numerically,
using the ansatz 
\beq
ds^2 = B^2(\xi)[-dt^2 + \exp(2 t)(dx^2 + dy^2)] + d\xi^2 + r^2(\xi) d\Omega_2^2 ,
\label{flatdSbrane}
\eeq
where the coordinates $x,y,t$ can be thought of as brane worldsheet
coordinates.   
Alternatively, one can use the closed de Sitter form of the worldsheet metric,
\beq
ds^2 = B^2(\xi)[-dt^2 + \cosh^2 t d{\Omega'}_2^2] + d\xi^2 + r^2(\xi)
d\Omega_2^2 , 
\label{closedbrane}
\eeq
with the same functions $B(\xi)$ and $r(\xi)$.  For a pure $dS_6$
space, these functions are given by 
\beq
B(\xi) =H_6^{-1} \cos (H_6 \xi) ,  ~~~~  r(\xi) = H_6^{-1}\sin (H_6 \xi) ,
\label{S6}
\eeq
with $\xi= \pi/2H_6$ corresponding to the de Sitter horizon.
(This metric covers only part of de Sitter space.)  

In the presence of a brane, the Einstein equations for $B(\xi)$ and
$r(\xi)$ with the ansatz (\ref{Fab}) for the Maxwell field have the
form 
\begin{eqnarray}
\label{inflating-brane-eqs}
 {1 \over {B(\xi)^2}} + {1\over {r(\xi)^2}} - {{B'(\xi)^2}\over {B(\xi)^2}}
-{{4 B'(\xi) r'(\xi)}\over {B(\xi) r(\xi)}} - {{r'(\xi)^2}\over
  {r(\xi)^2}} - {{2 B''(\xi)^2}\over {B(\xi)}} - {{2 r''(\xi)}\over
  {r(\xi)}} &=&  \Lambda + {{Q_m^2} \over {32\pi^2 r(\xi)^4}}~~,
\nonumber \\ \nonumber \\
 {{3}\over {B(\xi)^2}} + {1\over {r(\xi)^2}} - {{3
    B'(\xi)^2} \over {B(\xi)^2}} - {{6 B'(\xi) r'(\xi)}\over {B(\xi)
    r(\xi)}} - {{r'(\xi)^2}\over {r(\xi)^2}}&=&   \Lambda
+ {{Q_m^2} \over {32\pi^2 r(\xi)^4}}~~~~~~ \\ \nonumber \\
{{3}\over {B(\xi)^2}} - {{3 B'(\xi)^2} \over {B(\xi)^2}} - {{3 B'(\xi)
    r'(\xi)}\over {B(\xi) r(\xi)}} - {{3 B''(\xi)} \over {B(\xi)}} 
- {{r''(\xi)}\over
   {r(\xi)}}&=& \Lambda - {{Q_m^2} \over {32\pi^2
    r(\xi)^4}}~. \nonumber 
\end{eqnarray}
The first of these equations follows from the other two, so there are 
two independent equations for the two functions, $B(\xi)$ and $r(\xi)$.

Before discussing the solutions of these equations, let us consider a
Euclidean continuation of the metric (\ref{closedbrane}).  This can be
found by setting $t = i\theta - i\pi/2$, 
\beq
ds^2 = B^2(\xi) d{\Omega}_3^2 + d\xi^2 + r^2(\xi) d\Omega_2^2 .
\label{euclideandbrane}
\eeq
We are interested in compact, non-singular instanton solutions, so the
range of $\xi$ has to be finite, $0<\xi<\xi_m$, with the endpoints
corresponding to zeros of $B(\xi)$, 
\beq
B(0)=B(\xi_m)=0.
\label{bc1}
\eeq
These endpoints are non-singular, provided that
\beq
B'(0)=B'(\xi_m)=1
\label{bc2}
\eeq
and
\beq
r'(0)=r'(\xi_m)=0.
\label{bc3}
\eeq

The boundary conditions (\ref{bc1})-(\ref{bc3}) select a unique
solution of Eqs.~(\ref{inflating-brane-eqs}).  For specified values 
of $H_6$ and $Q_m$, the solution can be found numerically. For small 
values of $Q_m$,
\beq
Q_m\ll H_6^{-1}, 
\label{smallg}
\eeq
the brane horizon radius is much smaller than that of the cosmological
horizon at $\xi_m\approx \pi/2H_6$, and we expect the solution to be
well approximated by the spherical metric (\ref{S6}), except in the
vicinity of $\xi=0$.  We also expect that in this limit the brane
horizon radius is $r(0)\approx r_0$ with $r_0$ from (\ref{r0}) and the
brane tension is approximately given by (\ref{T2}).

\begin{figure}
\centering\leavevmode
\epsfysize=6cm \epsfbox{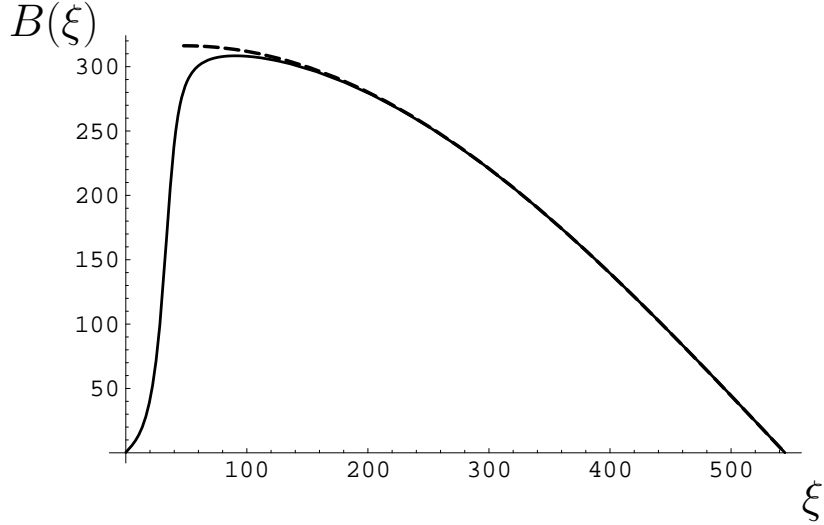}
\put(0,-10){\Large {\bf {$\xi$}}}
\put(-300,170){\Large {\bf {$B(\xi)$}}}
\caption[Fig 6] {Numerical solution for the function $B(\xi)$ for the
  magnetically charged inflating brane (solid line). We show for comparison the pure de
  Sitter solution Eq. (\ref{S6}) shifted to match the numerical
  solution at the cosmological horizon endpoint, namely $\xi_m$ (dashed
  line). We use here the parameters specified in Fig. 1 and the brane 
solution corresponds to the case $n=10$.}
\label{Bnumerical-1}
\end{figure}

\begin{figure}
\centering\leavevmode
\epsfysize=6cm \epsfbox{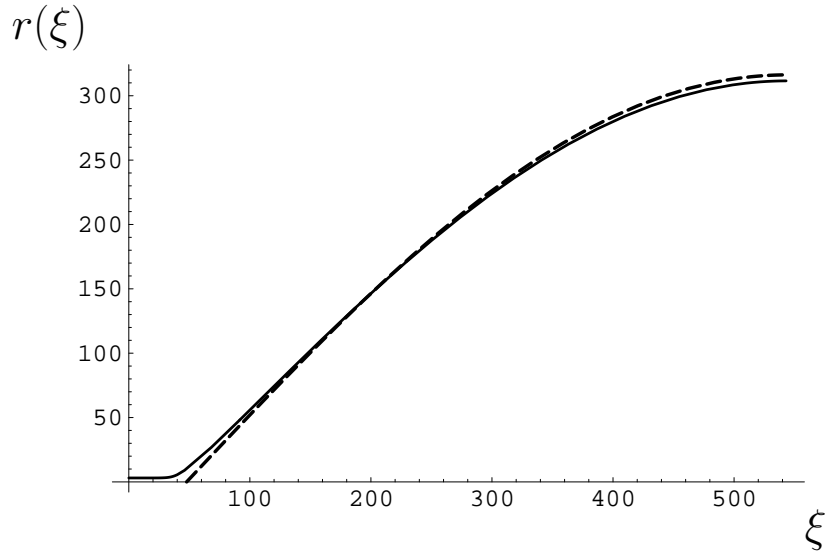}
\put(0,-10){\Large {\bf {$\xi$}}}
\put(-300,180){\Large {\bf {$r(\xi)$}}}
\caption[Fig 7] {Numerical solution for the function $r(\xi)$ (solid
  line) for the same parameters as in Figure \ref{Bnumerical-1}. The dashed line is the 
solution for $r(\xi)$ for the pure $dS_6$ case, Eq. (\ref{S6}). Note that we have shifted the pure de Sitter solution to make
the cosmological horizon coincide with $\xi_m$..}
\label{rnumerical-1}
\end{figure}

\begin{figure}
\centering\leavevmode
\epsfysize=6cm \epsfbox{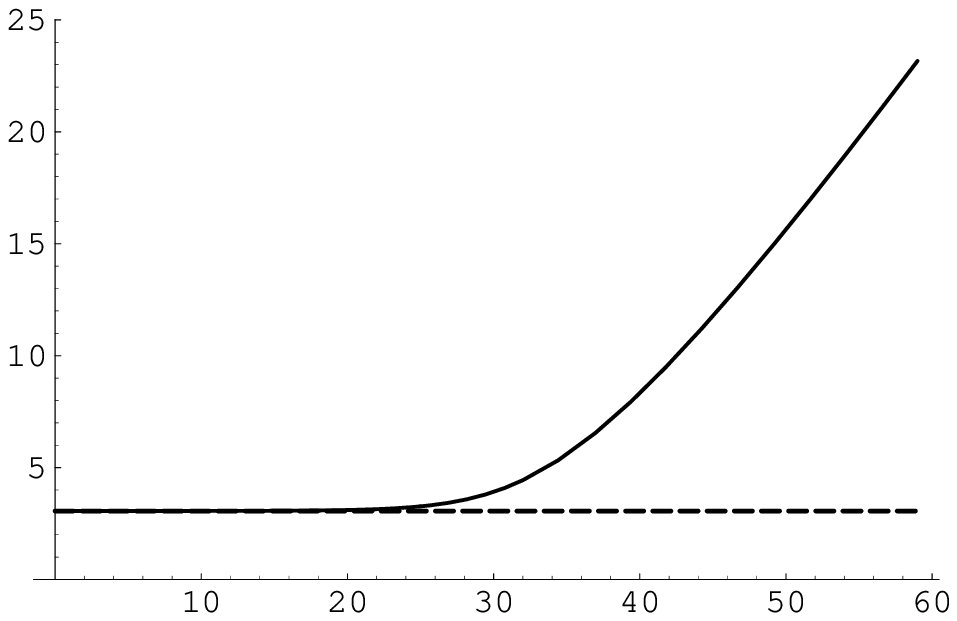}
\put(0,-10){\Large {\bf {$\xi$}}}
\put(-300,180){\Large {\bf {$r(\xi)$}}}
\caption[Fig 7] {Magnified plot of the numerical solution in
  Fig. \ref{rnumerical-1} for $r(\xi)$
  near the brane horizon (solid line). We also show for comparison the
analytic estimate of the brane horizon for this particular brane
solution (dashed line) given by Eq.~(\ref{r0}).}
\label{rnumerical-closeup}
\end{figure}

In Figs.~\ref{Bnumerical-1} and \ref{rnumerical-1} we find the
   numerical solution for
the case $e^2 = 2$, $\Lambda_6 = 10^{-4}$ and $n=10$. This value of
$n$ corresponds to a small magnetic charge so one can see from the figures 
that the functions only deviate slightly from their pure de Sitter 
counterparts. In Fig.~\ref{rnumerical-closeup} we show a close up of the
region near the black brane horizon and compare the asymptotic value
of $r$ with the analytic estimate in the small charge regime, namely,
$r_0$ from Eq. (\ref{r0}).

Analysis similar to that in Sec.~III.A indicates that the black brane
instanton can be pictured as a 2-brane, whose worldsheet has the form
of a 3-sphere and is wrapped around a ``big circle" of $S_6$, as
illustrated in Fig.~\ref{Alexfig1}.  As discussed in \cite{Basu}, such instantons
describe spontaneous nucleation of horizon-radius spherical branes in
de Sitter space.  The instanton action in the limit (\ref{smallg}) can
be estimated as   
\beq
S_{inst}\approx S_{dS_6} + A_3 H_6^{-3} T_2 ,
\label{Sapprox}
\eeq
where $A_3 = 2\pi^2$ is the volume of a unit 3-sphere and $T_2$
is the tension of the brane given by (\ref{T2}).   As for the black 
hole instanton, the presence of the brane does not induce any linear
in $T_2$ corrections to $S_{dS_6}$, because $dS_6$ is a minimum of the action.
The brane nucleation rate is given by 
\beq
\Gamma \sim \exp (-2\pi^2 H_6^{-3} T_2).
\eeq

\begin{figure}
\centering\leavevmode
\epsfysize=6cm \epsfbox{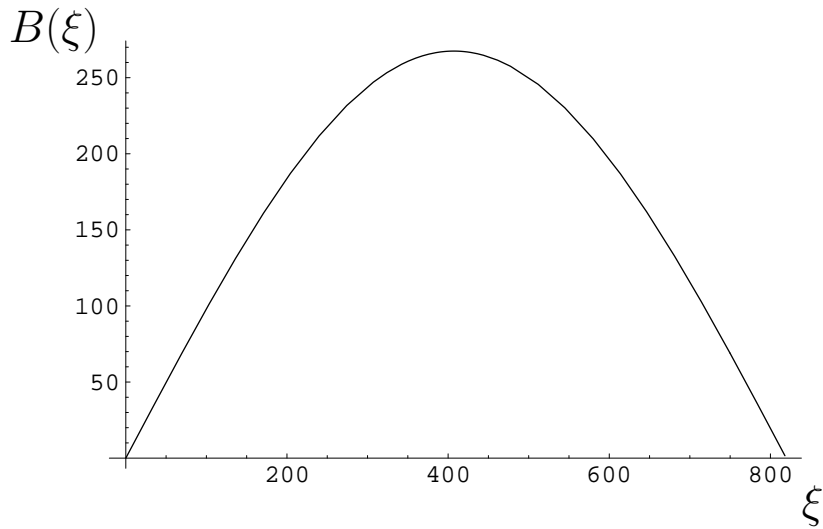}
\put(0,-10){\Large {\bf {$\xi$}}}
\put(-300,170){\Large {\bf {$B(\xi)$}}}
\caption[Fig 6] {Numerical solution for the function $B(\xi)$ for the
  magnetically charged inflating brane solution with $n=200$.}
\label{Bnumerical-2}
\end{figure}

\begin{figure}
\centering\leavevmode
\epsfysize=6cm \epsfbox{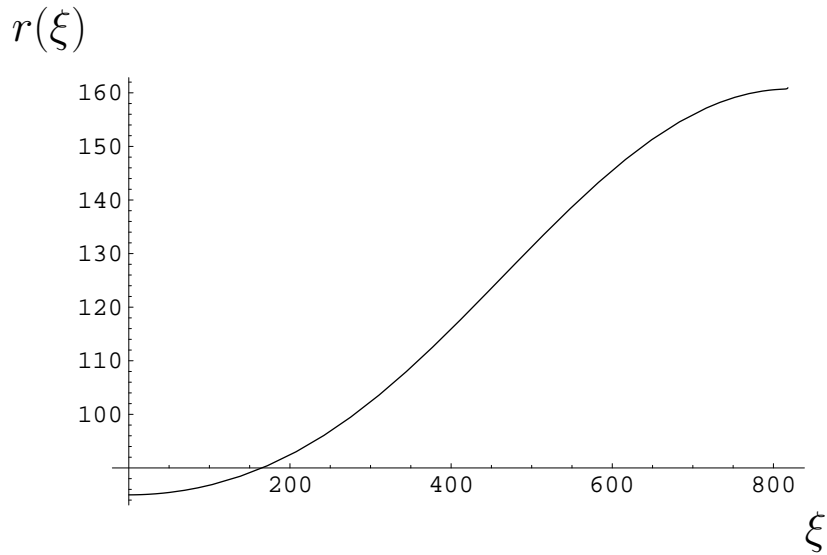}
\put(0,-10){\Large {\bf {$\xi$}}}
\put(-300,180){\Large {\bf {$r(\xi)$}}}
\caption[Fig 7] {Numerical solution for the function $r(\xi)$ for the
  case $n=200$.}
\label{rnumerical-2}
\end{figure}

We note that it follows from Eq.~(\ref{Lambda4}) that for small values
of $Q_m$ satisfying (\ref{smallg}) the $4d$ cosmological constant is
necessarily negative. For $Q_m\sim H_6^{-1}$, the brane and
cosmological horizons are comparable to one another, and the metric
significantly differs from (\ref{S6}) all the way to $r=r_c$.  In this
regime, the instanton action needs to be calculated numerically.  
An example in this regime is shown in Figs. \ref{Bnumerical-2} and \ref{rnumerical-2}.

The tunneling process can also be described using the $4d$ effective
theory. The instanton is then the usual Coleman-De Luccia (CdL) instanton. This
description was used by Carroll, Johnson and Randall (CJR)
\cite{Carroll09} and is equivalent to ours, but the geometry of the 
instanton and the resulting spacetime structure are not evident in this approach.
CJR performed a numerical calculation of the instanton action as a
function of the magnetic charge $Q_m$; the result is shown in their
Fig.~17. We have verified that their calculation agrees with our
approximate formula (\ref{Sapprox}). Since the graph obtained by CJR 
is nearly a straight line, our formula gives a good approximation in 
the entire range of $Q_m$.

CJR have also considered a Nariai-type instanton and compared its
action to that of the CdL-type instanton.  In the 
present case, the Nariai-type instanton is the Euclideanized 
unstable $dS_4\times S_2$ vacuum
solution, corresponding to the maximum of $V(\psi)$. Its action is
greater than that for the black brane instanton discussed above for
all values of $Q_m$, up to some critical value $Q_c^{(m)}$ at which
the two actions coincide\footnote{One can see how the inflating
    black brane solutions approach the Nariai limiting case by looking 
at the numerical solutions presented in Figs. \ref{Bnumerical-2} and 
\ref{rnumerical-2}. This is an intermediate solution where the 
function $B(\xi)$ approaches a pure sine function and
$r(\xi)$ has a small variation from $r_+$ to $r_c$ as one would expect
for a near Nariai solution.}. 
The black brane instanton does not exist 
for $Q_m>Q_c^{(m)}$, and no instanton solutions exist for
$Q_m>Q_{max}^{(m)}$ with $Q_{max}^{(m)}$ given by Eq.~(\ref{Qmax(m)}). The 
value of $Q_c^{(m)}$ depends non-trivially on $\Lambda_6$.

\subsection{Spacetime structure}

Let us now analyze the spacetime structure resulting from brane
nucleation.  The region between the brane horizon, $\xi=0$, and the
cosmological horizon, $\xi=\xi_m$, is described by the metric
(\ref{closedbrane}).  Surfaces of constant $\xi$ in this metric have
the geometry of $dS_3\times S_2$, indicating that the brane worldsheet
is effectively an expanding $3d$ de Sitter space.  In the vicinity of
the brane horizon at $\xi=0$, we have $B(\xi)\approx \xi$ and
$r(\xi)\approx r_+$, where $r_+ =r(0)$ is the brane horizon radius,   
\beq
ds^2 = \xi^2 (-dt^2 + \cosh^2t d{\Omega'}_2^2) + d\xi^2 + r_+^2 d\Omega_2^2 .
\label{nearhorizonbrane}
\eeq
Analytic continuation across the horizon is obtained by replacing
$\xi\to it$, $t\to \chi + i\pi/2$.  This gives 
\beq
ds^2 = -dt^2 + t^2 d{\cal H}_3^2 + r_+^2 d\Omega_2^2 ,
\label{nearhorizon2}
\eeq
where
\beq
d{\cal H}_3^2 = d\chi^2 + \sinh^2\chi d\Omega^2
\eeq
is a $3d$ hyperbolic metric of unit curvature radius.  The metric
(\ref{nearhorizon2}) describes an expanding, open $4d$ FRW universe
with 2 extra dimensions compactified on a sphere.

\begin{figure}
\centering\leavevmode
\epsfig{file=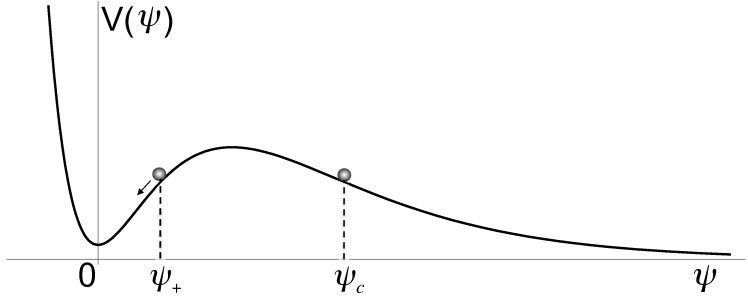,width=5cm,angle=-90}
\caption {Scalar field potential of the effective $4d$ theory. 
In the newly formed compactified region, $\psi$ starts at $\psi_+$ 
and rolls to $\psi=0$.}
\label{Alexfig3}
\end{figure}

The metric (\ref{nearhorizon2}) applies only in the vicinity of $t=0$,
where the $4d$ universe is curvature-dominated.  More generally, the
metric behind the black brane horizon has the form 
\beq
ds^2 = -dt^2 + a^2(t) d{\cal H}_3^2 + r^2(t) d\Omega_2^2.
\label{dS4xS2}
\eeq
The evolution of $a(t)$ and $r(t)$ can be found by numerically solving
the Lorentzian version of the $6d$ Einstein equations. 
Alternatively, we can use the effective $4d$ theory
(\ref{4Deffectiveaction}).  The $4d$ metric is given by 
\beq
ds_4^2 = -d{\tilde t}^2 + {\tilde a}^2({\tilde t}) d{\cal H}_3^2
\eeq
and the scalar field $\psi$ is a function of ${\tilde t}$ only.  The
$4d$ and $6d$ descriptions are connected by the relations 
\beq
dt = e^{-\psi({\tilde t})/2M_4}d{\tilde t},
\eeq
\beq
a(t) = e^{-\psi({\tilde t})/2M_4} {\tilde a}({\tilde t}),
\eeq
\beq
r(t) = R e^{\psi({\tilde t})/2M_4} .
\eeq
The $4d$ evolution equations have the form
\beq
{\ddot \psi}+3{{\dot {\tilde a}}\over{\tilde a}} {\dot\psi} + V'(\psi) = 0 ,
\label{e1}
\eeq
\beq
\left({{\dot {\tilde a}}\over{\tilde a}}\right)^2 -{1\over{\tilde
    a}^2} = {1\over{3M_4^2}} \left({{\dot\psi}^2\over{2}} + V(\psi)\right),
\label{e2}
\eeq
where dots stand for derivatives with respect to ${\tilde t}$ and 
$V(\psi)$ given by (\ref{V}).  

\begin{figure}
\centering\leavevmode
\epsfig{file=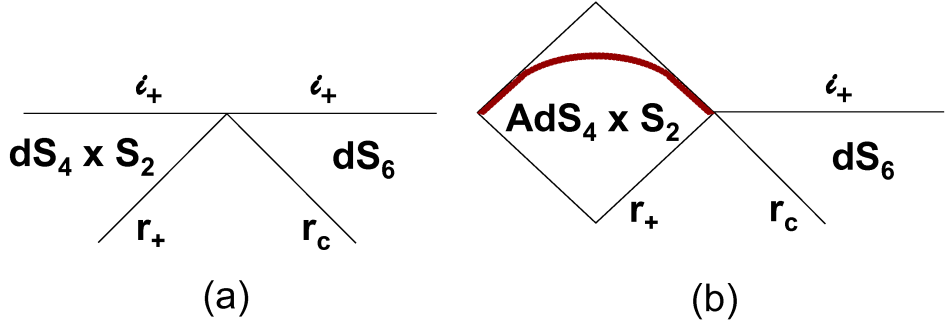,width=5cm,angle=-90}
\caption {Spacetime diagrams for transitions from  $dS_6$ to 
$dS_4\times S_2$ (a) and $AdS_4\times S_2$ (b).}
\label{Alexfig4}
\end{figure}

From a $4d$ point of view, our tunneling process is the usual
Coleman-De Luccia (CdL) tunneling through the barrier in the potential
$V(\psi)$.  The field values $\psi_c$ and $\psi_+$, corresponding
respectively to $r_c$ and $r_+$ are located on the opposite sides of 
the barrier (see Fig.~\ref{Alexfig3}).  The evolution starts at ${\tilde t}=0$ with 
$\psi=\psi_+$ and ${\tilde a}=0$, and the field $\psi$ starts rolling 
towards the potential minimum at $\psi = 0$.  The long-term evolution 
depends on whether this minimum has positive or negative energy
density. For $\Lambda_4 <0$, we have an $AdS_4\times S_2$ vacuum, and 
the evolution ends in a singular big crunch, while for $\Lambda_4 >0$, 
the evolution is non-singular and the metric asymptotically approaches 
a $4d$ de Sitter space with $\psi\to 0$.  The corresponding spacetime
diagrams are shown in Fig.~\ref{Alexfig4}.

Nariai-type instantons are similar to Hawking-Moss instantons
\cite{HawkingMoss}. As discussed in \cite{Carroll09}, they describe 
nucleation of regions of size comparable to the horizon, filled with 
the unstable vacuum at the maximum of $V(\psi)$. Small perturbations 
will cause $\psi$ to roll away from the maximum. As a result, the
newly formed region will either disintegrate into $dS_6$ and $(A)dS_4
\times S_2$ domains, or, if the potential $V(\psi)$ is sufficiently 
flat near the maximum, it will become a site of eternal inflation, with 
$dS_6$ and $(A)dS_4 \times S_2$ sub-regions constantly being formed.

\section{From $dS_4\times S_2$ to $dS_6$}

The decompactification transition from $dS_4\times S_2$ to $dS_6$ has
been discussed in Refs.~\cite{BPSPV09,Carroll09}.  In terms of the effective
$4D$ theory, this is the usual false vacuum decay.  A bubble with
$\psi\sim \psi_c$ nucleates in the inflating background of $\psi=0$
vacuum.  As the bubble expands, the field rolls off to infinity in the
bubble interior.  The tunneling is described by the same CdL instanton
as for the inverse transition, $dS_6\to dS_4\times S_2$, and the tunneling
rate is 
\beq 
\Gamma \sim \exp(S_0-S_{inst}) ,
\label{GammaDC}
\eeq
where 
\beq
S_0 = -24\pi^2 M_4^4/\Lambda_4 ,
\eeq
with $M_4$ and $\Lambda_4$ from Eqs.~(\ref{M4}) and (\ref{Lambda4}),
respectively.  The rate (\ref{GammaDC}) has been estimated in
\cite{BPSPV09} and has been calculated numerically for different
values of parameters in Ref.~\cite{Carroll09}.

\begin{figure}
\centering\leavevmode
\epsfig{file=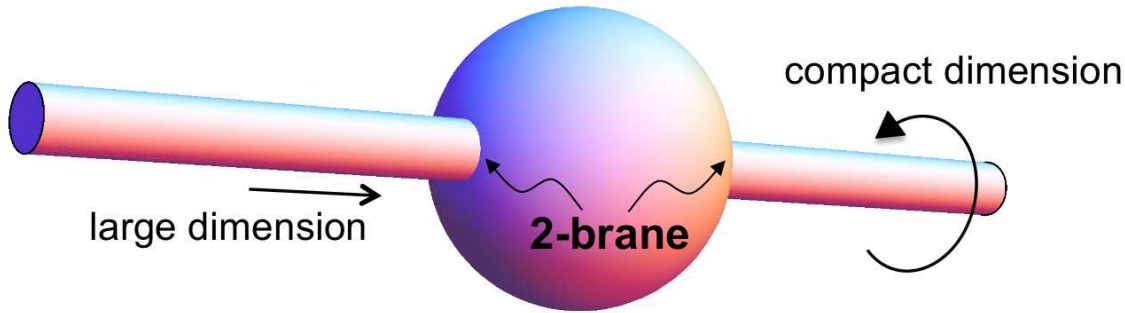,width=16cm}
\caption {A spatial slice through the decompactification bubble
  spacetime. The junctions marked as ``2-branes'' in this
  lower-dimensional analogue become a $2$-sphere in $5d$.}
\label{Alexfig5}
\end{figure}

The $4d$ evolution after tunneling is described by the same
equations (\ref{e1}), (\ref{e2}) as for the inverse transition, except 
now the evolution starts with $\psi({\tilde t}=0)=\psi_c$ and the
field $\psi$ rolls towards $\psi=\infty$, so the extra dimensions
become large. The role of the bubble wall is played by the expanding 
black 2-brane. A spacelike slice through the decompactification bubble 
spacetime is illustrated in Fig.~\ref{Alexfig5}.

The 6-dimensional metric in the bubble interior has the form 
(\ref{dS4xS2}). At late times, $t\to\infty$, the metric approaches
that of $dS_6$ \cite{Decompactification}, with $a(t)\approx H_6^{-1}
\sinh(H_6 t)$ and $r(t)\approx H_6^{-1}\cosh (H_6 t)$. This is a
somewhat unfamiliar, ``mixed" slicing of de Sitter space, with open
slices in 3 out of 5 spatial dimensions and closed spherical slices 
in the remaining two directions. The compact dimensions always remain 
compact, but asymptotically become very large, and the metric becomes 
locally isotropic in the limit.

\section{Conclusions}

The Einstein-Maxwell theory in 6 dimensions admits a rich landscape of
vacua and provides a useful toy model for investigating the dynamics
of the multiverse.  The landscape of this theory includes a $dS_6$
vacuum, a number of $dS_4$ and $AdS_4$ vacua with extra dimensions
compactified on $S_2$, and a number of $AdS_2$ vacua with extra
dimensions compactified on $S_4$. There are also some perturbatively 
unstable $dS_4\times S_2$ and $dS_2\times S_4$ vacua.  In this paper we studied quantum
tunneling transitions between vacua of different effective
dimensionality.  We identified the appropriate instantons and
described the spacetime structure resulting from the tunneling. Our 
results reinforce and extend the earlier analyses in
Refs.~\cite{Carroll09} and \cite{BPSPV09}.

We found that compactification transitions from $dS_6$ to $AdS_2
\times S_4$ occur through nucleation of pairs of electrically charged
black holes.  The charge of these black holes is quantized in units of
the elementary charge of the theory, and their mass is determined by
the condition that the temperature at the black hole horizon $(r=r_+)$
is the same as that at the cosmological horizon $(r=r_c)$ in the
corresponding Reissner-Nordstrom-de Sitter solution.  These black
holes can be thought of as 0-branes of the theory.  In the limit when
$r_+ \ll r_c$ the black holes become nearly extremal.  The instanton
in this limit can be pictured as $S_6$ (Euclideanized de Sitter) with
a 0-brane worldline running along a big circle.

Transitions from $dS_6$ to $dS_4 \times S_2$ and $AdS_4 \times S_2$
proceed through nucleation of spherical, magnetically charged black
2-branes.  Once again, in the limit when the black brane horizon
radius $r_+$ is small compared to $r_c$, the corresponding instanton
can be pictured as a brane whose Euclidean worldsheet (which has the form
of a 3-sphere) is wrapped around a ``big circle'' of $S_6$.  The process
of black brane formation through quantum tunneling is very similar to
nucleation of spherical domain walls during inflation, which was
analyzed in Ref.~\cite{Basu}.

The initial radius of a nucleating brane is set by the de Sitter
horizon, $H_6^{-1}$.  Once the brane is formed, this radius is
stretched by the exponential expansion of the universe, while the
transverse dimension of the brane (which can be identified with its
horizon radius $r_+$) remains fixed.  Behind the horizon, in the black 
brane interior, the spacetime is effectively 4-dimensional, with the 
extra two dimensions compactified on $S_2$. Observers in this newly 
formed compactified region see only a homogeneous FRW universe, 
approaching either dS or AdS space. 

Transitions from $dS_6$ to unstable $dS_2\times S_4$ (or $dS_4\times
S_2$) vacua are mediated by Nariai-type instantons, which are similar 
to Hawking-Moss instantons.  Small perturbations cause fragmentation
of these vacua into regions of $AdS_2\times S_4$ (or $AdS_4\times S_2$) and $dS_6$.

We also discussed decompactification transitions $dS_4 \times S_2 \to
dS_6$, in which the effective dimension of the daughter vacuum is
higher than that of the parent vacuum.  These transitions are
described by the same instanton as the inverse process, $dS_6 \to dS_4
\times S_2$, but the resulting spacetime structure is rather
different.   Observers in the parent vacuum see
nucleation and subsequent expansion of spherical bubbles, as in
ordinary CdL vacuum decay.  The role of the bubble wall is played by a 
spherical magnetically charged black brane (the same kind of brane as 
in the inverse transition). The bubble interior is initially
anisotropic, but as the compact dimensions expand, it approaches 
local isotropy, with the metric approaching the $6d$ de Sitter space. 
However, the anisotropy may still be observable in the CMB
if inflation inside the bubble terminates after a 
relatively small number of $e$-foldings. These issues will be
discussed separately in \cite{BlancoPillado:2010uw}.

Apart from the topology-changing transitions that we discussed here, 
our model admits tunneling transitions between $(A)dS_4 \times S_2$
vacua having the same topology, but characterized by different values of the
magnetic flux. Such tunnelings, which are analogous to the Schwinger 
process, and the corresponding instantons, have been discussed in
\cite{BPSPV09}. The role of the bubble walls in the resulting bubbles is played
by magnetically charged branes, which can be either solitonic branes 
(if the model is augmented by adding the appropriate Higgs fields) 
or black branes of the type described here. In the latter case, the region inside
the black brane horizon is a new $(A)dS_4 \times S_2$ universe. Thus, 
the tunneling results in the formation of two compactified regions: 
one in the interior of the expanding bubble, where the flux has been changed with
respect to the parent vacuum, and the other in the interior of the black brane itself.

\begin{figure}
\centering\leavevmode
\epsfig{file=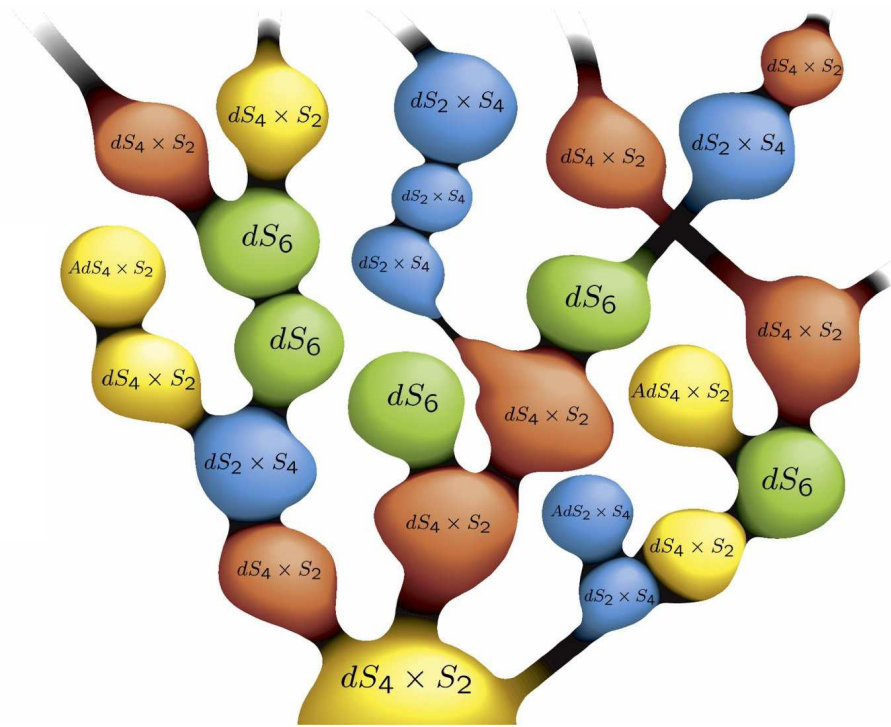,width=10cm,angle=-90}
\caption {A spacelike slice through a multi-dimensional multiverse. 
The dark regions represent black holes and black branes. This is an 
adaptation of Andrei Linde's picture in \cite{Linde-website}.}
\label{cactus}
\end{figure}

Topology-changing transitions of the kind we discussed in this paper 
will inevitably occur in the course of eternal inflation.  As a
result, the multiverse will be populated by vacua of all possible 
dimensionalities.  A spacelike slice through such a multiverse might 
look like Linde's famous ``cactus" picture \cite{Linde-website}; 
its version adapted for our present model is shown in 
Fig.~\ref{cactus}.  We shall discuss the structure of this 
multi-dimensional multiverse and its implications for the 
measure problem in a separate paper.

Finally, as Fig.~\ref{cactus} suggests, we expect that there should be 
other kinds of instantons that interpolate directly between the 
$dS_4 \times S_2$ and $AdS_2 \times S_4$ sectors in our model. These 
would be somewhat more complicated solutions that will not only change 
the topology of spacetime, but would also act as sources
and sinks for the different fluxes involved.

\section{Acknowledgments}

We are grateful to Raphael Bousso, Oscar Dias, Jaume Garriga, Andrei Linde, Maulik Parikh, Oriol
Pujolas, Mike Salem and Ben Shlaer for very helpful
discussions. J.J. B.- P. would like to thank the Theory division at CERN for
their support and hospitality while part of this work was being completed. This work 
was supported in part by the National Science Foundation Grants
06533561 (J.J.B-P.) and 0353314 (A.V.).

\appendix

\section{The Electric Sector in the four form language}\label{appelecsec}

In this appendix we will discuss $6$ dimensional Einstein gravity coupled to a 
$4$-form field. We will call this the electric sector 
of the $6d$ model. In this case we have $2$ large dimensions and 
$4$ compact ones. The Lagrangian is given by
\beq
S=\int{d^6 x \sqrt{-\tilde g} \left( {{1}\over 2}
{\tilde R}^{(6)} - {1\over {48}} F_{MNPQ}F^{MNPQ} - \Lambda_6\right)},
\label{EM-6D-action-4-form}
\eeq
which is the dual theory of the electromagnetic formulation given in
the main part of the text. The corresponding field equations are

\beq
{\tilde R}_{MN}^{(6)} - {1\over 2} {\tilde g}_{MN} {\tilde R}^{(6)} 
= T_{MN}
\eeq
and
\beq
{1\over{\sqrt{-\tilde g}}} \partial_M \left(\sqrt{-\tilde g} F^{MNPQ}\right)=0,
\eeq
with the energy-momentum tensor given by

\beq
T_{MN} ={{1} \over {4!}} \left(4  F_{MPQR} F_{N}^{~PQR} - {1\over 2} 
{\tilde g}_{MN} F^2 \right) - {\tilde g}_{MN}  \Lambda_6~.
\eeq 
 
We will look for solutions of this model with the spacetime metric
given by a two-dimensional maximally symmetric space of constant
curvature,\footnote{As before, $H^2$ can be positive or negative,
depending on whether we are talking about de Sitter or anti-de Sitter
spaces.} $R^{(2)} = 2 H^2$, and a static extra-dimensional 4-sphere
of fixed radius, namely a metric of the form
\beq
ds^2= {\tilde g}_{MN} dx^M dx^N = {\tilde g}_{\mu \nu} d x^{\mu}
d x^{\nu} + R^2 d\Omega_4^2~.
\label{6D-metricelectric}
\eeq 

With this ansatz, we obtain the following components of the $6d$
Einstein tensor:
\begin{eqnarray}
{\tilde G}_{\mu \nu}^{(6)} 
&=& - {6\over {R^2}} \tilde g_{\mu \nu}~,\\
{\tilde G}_{ab}^{(6)} &=&  -\left(H^2 + {3\over {R^2}}\right){\tilde g_{ab}} ~,
\end{eqnarray}
where we have used $\mu$ and $\nu$ to denote the 
two dimensional coordinates and $a$ and $b$ run over the
four extra dimensions on the sphere.

The field strength is taken to be proportional to the volume form on
$S_q$, namely,

\beq
F_{\theta \phi\psi\chi} = {Q_e\over{A_4}}~({\rm det} g_{ab})^{1/2} 
= {Q_e\over{A_4}}~\sin^3{\theta}\sin^2{\phi}\sin{\psi},
\eeq
which gives rise to the following energy-momentum tensor

\beq \label{Tmunuhdsol}
T_{\mu \nu} = - {\tilde g_{\mu \nu}} \left({{Q_e^2}\over{2 A_4^2 R^8}} 
+ \Lambda_6\right)
\eeq
and
\beq \label{Tijhdsol}
T_{ab} = {\tilde g}_{ab} \left({{Q_e^2}\over{2 A_4^2 R^8}} - \Lambda_6\right),
\eeq
where, as before,
\beq
Q_e = n e.
\eeq

Equating the Einstein tensor and the stress-energy tensor we find
\beq
H^2 = {\Lambda_6 \over {2}}\left(1- {3{Q_e^2}\over{2 A_4^2 \Lambda_6 R^8 }}\right)\label{quarticinH}
\eeq
and
\beq
 {3 \over {R^2}} = {\Lambda_6 \over {2}}\left(1+{{Q_e^2}\over{2 A_4^2 \Lambda_6 R^8}}
     \right) \label{quarticinR}.
\eeq

These equations are quartic in $R^2$.  Upon solving Eq. (\ref{quarticinR}) for $R^2$
we find that two of the four solutions yield negative $R^2$, and are
thus discarded. The solutions are algebraically complicated, and are 
analyzed in Appendix \ref{quarticequation}.  The key point for this
study is that for the two viable solutions it can be shown that upon 
substituting into Eq.~(\ref{quarticinH}) one solution always yields 
a positive $H^2$ while the other solution always yields a negative $H^2$.

\section{Solving for $R^2$}\label{quarticequation}

We need to solve Eq.~(\ref{R^2}),
\beq
 {3 \over {R^2}} = {\Lambda_6 \over {2 }}\left(1+{{Q_e^2}\over{2 A_4^2
     \Lambda_6 R^8}}
     \right) . 
\label{eqR}
\eeq

Multiplying this equation by $(Q_e/A_4)^{2/3}$ and introducing a new
variable $x = R/(Q_e/A_4)^{1/3}$, we have
\beq
{3\over{x^2}}={{\Lambda_6 (Q_e/A_4)^{2/3}}\over{2}} + {1\over{4x^8}},
\eeq
or
\beq
1-Kx^2 = {{1}\over {12x^6}},
\label{Kx}
\eeq
where 
\beq
K = {{(Q_e/A_4)^{2/3}\Lambda_6}\over {6}}.
\label{K}
\eeq

The function on the l.h.s. of eq. (\ref{Kx}) is convex and the 
function on r.h.s. is concave.  For sufficiently small values of $K$, 
there are two roots.  As $K$ is increased, the roots approach one 
another until they coincide at some critical value $K_{max}$ 
(corresponding to some flux quantum number $n_{max}$).  
For $K>K_{max}$, there are no solutions.

The critical value $K_{max}$ can be found by noticing that at this 
value of $K$ the graphs of the functions on the two sides of
eq.(\ref{Kx}) are tangent to one another at the point where they
meet. This yields the condition
\beq
x^8 = {{1}\over{4K}}.
\eeq
Combining this with Eq. (\ref{Kx}), we find
\beq
K_{max} ={{3^{4/3}}\over{4}} .
\eeq
Using Eqs.~(\ref{K}), (\ref{Qe}) and (\ref{H2}), it can be verified
that the corresponding values of $Q_e$, $n$ and $H$ are
\beq
Q_{max}^{(e)} = 9A_4 \left({{3}\over {2\Lambda_6}}\right)^{3/2} , 
\label{Qc}
\eeq
\beq
n_{max} = {{9A_4}\over{e}}\left({3\over{2\Lambda_6}}\right)^{3/2} ,
\eeq
and $H = 0$.  Since the two solutions of (\ref{eqR}) merge at $H =0$, 
it is clear that for $K<K_{max}$ one of these solutions has $H>0$ and 
the other $H<0$.

For $K\ll K_{max}$, the solutions of (\ref{Kx}) can be approximated as
\beq
x_1\approx 12^{-1/6}, ~~~~~~~  x_2 \approx K^{-1/2}.
\eeq
(Note that in this case $K\ll 1$.)

\section{Inflating brane equations of motion}\label{braneeom}

Taking into account the following ansatz for the metric,

\beq
ds^2 = B^2(\xi)[-dt^2 + \cosh^2 t d{\Omega'}_2^2] + d\xi^2 + r^2(\xi)
d\Omega_2^2 , 
\eeq

we obtain an Einstein tensor of the form,

\begin{eqnarray}
\label{eq:Einstein}
G^{t}_{t}   &=& G^{\theta'}_{\theta'} = G^{\phi'}_{\phi'} = - {1 \over
  {B(\xi)^2}} - {1\over {r(\xi)^2}} + {{B'(\xi)^2}\over {B(\xi)^2}}
+{{4 B'(\xi) r'(\xi)}\over {B(\xi) r(\xi)}} + {{r'(\xi)^2}\over
  {r(\xi)^2}} + {{2 B''(\xi)^2}\over {B(\xi)}} + {{2 r''(\xi)}\over {r(\xi)}}~,
\nonumber \\ \nonumber \\
G^{\xi}_{\xi} &=& - {{3}\over {B(\xi)^2}} - {1\over {r(\xi)^2}} + {{3
    B'(\xi)^2} \over {B(\xi)^2}} + {{6 B'(\xi) r'(\xi)}\over {B(\xi)
    r(\xi)}} + {{r'(\xi)^2}\over {r(\xi)^2}}~,\nonumber\\ \nonumber \\
G^{\theta}_{\theta} &=& G^{\phi}_{\phi} = - {{3}\over {B(\xi)^2}} + {{3
    B'(\xi)^2} \over {B(\xi)^2}} + {{3 B'(\xi) r'(\xi)}\over {B(\xi)
    r(\xi)}} + {{3 B''(\xi)} \over {B(\xi)}} + {{r''(\xi)}\over
  {r(\xi)}}~. \nonumber
 \end{eqnarray}

On the other hand, the energy-momentum tensor of the brane in this metric
can be computed to be,
\begin{eqnarray}
\label{eq:emtensor}
T^{t}_{t}   &=& T^{\theta'}_{\theta'} = T^{\phi'}_{\phi'} = - \Lambda_6
- {{Q_m^2} \over {32\pi^2 r(\xi)^4}}
\nonumber\\
T^{\xi}_{\xi} &=&  - \Lambda_6
- {{Q_m^2} \over {32\pi^2 r(\xi)^4}}
\nonumber\\
T^{\theta}_{\theta} &=& T^{\phi}_{\phi} =  - \Lambda_6
+{{Q_m^2} \over {32\pi^2 r(\xi)^4}} \nonumber
\end{eqnarray}

We require that our solution must be free of singularities at
the brane horizon as well as the cosmological horizon, which in
turn means that the most general expansion around those points
should be of the form
\begin{eqnarray}
\label{eq:horizon-expansion}
B(\xi) &=& \xi + B_3~\xi^3 + ... \nonumber\\
r(\xi) &=& r_H + r_2~\xi^2 + ...
\end{eqnarray}
where we have assumed that the position of the horizon in question
is at $\xi = 0$ and its radius is $r_H$. Plugging this expansion back
into the equations of motion we arrive at the following expression for
the coefficients $B_3$ and $r_2$ in terms of $\Lambda_6$, $Q_m$ and $r_H$, namely,

\begin{eqnarray}
B_3 &=& - {{4 r_H^2 - 5 (Q_m/4\pi)^2 + 2 r_H^4 \Lambda_6}\over
  {144 r_H^4}} \nonumber\\
r_2 &=& {{4 r_H^2 - 3 (Q_m/4\pi)^2 - 2 r_H^4  \Lambda_6}\over
  {32 r_H^3}}~.
\end{eqnarray}

Using this expansion we can now integrate the equations of motion
forward in $\xi$ up to the other horizon. The value of $r_H$ is then fixed 
by imposing the regularity condition on the other horizon.


\begin{thebibliography}{99}

\bibitem{Susskind}
  L.~Susskind,
  ``The anthropic landscape of string theory,'' (2003), arXiv:hep-th/0302219.
  

\bibitem{CdL}
  S.~R.~Coleman and F.~De Luccia,
  ``Gravitational Effects On And Of Vacuum Decay,''
  Phys.\ Rev.\  D {\bf 21}, 3305 (1980).
  

\bibitem{AV83}
  A.~Vilenkin,
  ``The Birth Of Inflationary Universes,''
  Phys.\ Rev.\  D {\bf 27} (1983) 2848.
  

\bibitem{Linde86}
  A.~D.~Linde,
  ``Eternally Existing Selfreproducing Chaotic Inflationary Universe,''
  Phys.\ Lett.\  B {\bf 175}, 395 (1986).

\bibitem{Linde88}
A. D. Linde and M. I. Zelnikov, 
``Inflationary universe with fluctuating dimension,'' Phys. 
Lett. B 215, 59 (1988). 

\bibitem{Decompactification}
  S.~B.~Giddings and R.~C.~Myers,
  ``Spontaneous decompactification,''
  Phys.\ Rev.\  D {\bf 70}, 046005 (2004).

\bibitem{BPSPV09}
  J.~J.~Blanco-Pillado, D.~Schwartz-Perlov and A.~Vilenkin,
  ``Quantum Tunneling in Flux Compactifications,''
  JCAP {\bf 0912}, 006 (2009)
  [arXiv:0904.3106 [hep-th]].

\bibitem{Carroll09}
S.~M.~Carroll, M.~C.~Johnson and L.~Randall,
  ``Dynamical compactification from de Sitter space,''
  JHEP {\bf 0911}, 094 (2009)
 [arXiv:0904.3115 [hep-th]].



\bibitem{Freund-Rubin}
  P.~G.~O.~Freund and M.~A.~Rubin,
  ``Dynamics Of Dimensional Reduction,''
  Phys.\ Lett.\  B {\bf 97}, 233 (1980).


\bibitem{EM6D}
  S.~Randjbar-Daemi, A.~Salam and J.~A.~Strathdee,
  ``Spontaneous Compactification In Six-Dimensional Einstein-Maxwell Theory,''
  Nucl.\ Phys.\  B {\bf 214}, 491 (1983).
 


\bibitem{flux-compactifications}
  M.~R.~Douglas and S.~Kachru,
  ``Flux compactification,''
  Rev.\ Mod.\ Phys.\  {\bf 79}, 733 (2007).

\bibitem{Bousso:2002fi}
 R.~Bousso, O.~DeWolfe and R.~C.~Myers,
 ``Unbounded entropy in spacetimes with positive cosmological constant,''
 Found.\ Phys.\  {\bf 33}, 297 (2003)
 [arXiv:hep-th/0205080].


\bibitem{Krishnan:2005su}
 C.~Krishnan, S.~Paban and M.~Zanic,
 ``Evolution of gravitationally unstable de Sitter compactifications,''
 JHEP {\bf 0505}, 045 (2005)
 [arXiv:hep-th/0503025].

\bibitem{Gibbons95}
  G.~W.~Gibbons, G.~T.~Horowitz and P.~K.~Townsend,
  ``Higher Dimensional Resolution Of Dilatonic Black Hole Singularities,''
  Class.\ Quant.\ Grav.\  {\bf 12} (1995) 297.


\bibitem{Gregory}
  R.~Gregory,
  ``Cosmic p-Branes,''
  Nucl.\ Phys.\  B {\bf 467}, 159 (1996).



\bibitem{Tangherlini}
  F.~R.~Tangherlini,
  ``Schwarzschild field in n dimensions and the dimensionality of space
  problem,''
  Nuovo Cim.\  {\bf 27}, 636 (1963).



\bibitem{Moss}
  F.~Mellor and I.~Moss,
  ``Black Holes And Quantum Wormholes,''
  Phys.\ Lett.\  B {\bf 222}, 361 (1989);
  F.~Mellor and I.~Moss,
  ``Black Holes and Gravitational Instantons,''
  Class.\ Quant.\ Grav.\  {\bf 6}, 1379 (1989).


\bibitem{Romans}
  L.~J.~Romans,
  ``Supersymmetric, cold and lukewarm black holes in cosmological
  Einstein-Maxwell theory,''
  Nucl.\ Phys.\  B {\bf 383}, 395 (1992).
  
\bibitem{Mann}
  R.~B.~Mann and S.~F.~Ross,
  ``Cosmological production of charged black hole pairs,''
  Phys.\ Rev.\  D {\bf 52}, 2254 (1995).



\bibitem{Dias}
  O.~J.~C.~Dias and J.~P.~S.~Lemos,
  ``Pair creation of higher dimensional black holes on a de Sitter
  background,''
  Phys.\ Rev.\  D {\bf 70}, 124023 (2004); See also hep-th/0410279 v3.




\bibitem{MyersPerry86}
  R.~C.~Myers and M.~J.~Perry,
  ``Black Holes In Higher Dimensional Space-Times,''
  Annals Phys.\  {\bf 172}, 304 (1986).


\bibitem{Basu}
  R.~Basu, A.~H.~Guth and A.~Vilenkin,
  ``Quantum creation of topological defects during inflation,''
  Phys.\ Rev.\  D {\bf 44}, 340 (1991);
  R.~Basu and A.~Vilenkin,
  ``Nucleation of thick topological defects during inflation,''
  Phys.\ Rev.\  D {\bf 46} (1992) 2345.

\bibitem{Parikh}
M.~Parikh,
 ``Enhanced instability of de Sitter space in Einstein-Gauss-Bonnet
gravity,''  arXiv:hep-th/0909.3307.


\bibitem{Bousso1}
R.~Bousso and S.~W.~Hawking, 
``Pair creation of black holes during inflation,''
Phys.\ Rev.\  D {\bf 54}, 6312 (1996).


\bibitem{Bousso2}
  R.~Bousso,
  ``Quantum global structure of de Sitter space,''
  Phys.\ Rev.\  D {\bf 60}, 063503 (1999).


\bibitem{Lu}
  J.~X.~Lu,
  ``ADM masses for black strings and p-branes,''
  Phys.\ Lett.\  B {\bf 313}, 29 (1993).
  
\bibitem{HawkingMoss}
  S.~W.~Hawking and I.~G.~Moss,
  ``Supercooled Phase Transitions In The Very Early Universe,''
  Phys.\ Lett.\  B {\bf 110}, 35 (1982).

 \bibitem{BlancoPillado:2010uw}
 J.~J.~Blanco-Pillado and M.~P.~Salem,
 ``Observable effects of anisotropic bubble nucleation,''
 arXiv:1003.0663 [hep-th].

\bibitem{Linde-website}
A.~D.~Linde, ``The Self-Reproducing inflationary universe,''  Scientific American  {\bf 271}, 32 (1994).


  

\end{thebibliography}
\end{document}